\documentclass{ws-ijmp}
\usepackage{hyperref}

\begin{document}

\markboth{A. A. Grib and Yu. V. Pavlov}
{Particles with Negative Energies in Black Holes}

\title{\uppercase{Particles with negative energies in black holes}}

\author{\footnotesize A. A. GRIB\footnote{
Present address:
Theoretical Physics and Astronomy Department, The Herzen  University,
48, Moika, St.\,Petersburg, 191186, Russia}}

\address{
A. Friedmann Laboratory for Theoretical Physics,\\
30/32 Griboedov can., St.\,Petersburg, 191023, Russia\\
andrei\_grib@mail.ru}

\author{YU. V. PAVLOV}

\address{
Institute of Mechanical Engineering,\\
Russian Academy of Sciences,\\
61 Bolshoy, V.O., St.\,Petersburg, 199178, Russia\\
yuri.pavlov@mail.ru}

\maketitle

\vspace{11pt}
\begin{abstract}
    The problem of the existence of particles with negative energies
inside and outside of Schwarzschild, charged and rotating black holes
is investigated.
    Different  definitions of the energy of the particle inside
the Schwarzschild black hole are analyzed and
it is shown in what cases this energy can be negative.
    A comparison is made for the cases of rotating black holes described
by the Kerr metric when the energy of the particle can be negative in
the ergosphere and the Reissner--Nordstr{\o}m metric.
\end{abstract}


\vspace{11pt}
{\section{Introduction}
\label{secUFNIntr}
}
\noindent
    The problem of time in cosmology, mainly in quantum cosmology, is one
of the favorite topics of Prof. M.~Castagnino.\cite{Castagnino89,Castagnino98}
    In this paper, being a tribute to his anniversary, we shall discuss
a similar problem arising inside and outside black holes.
    It is the problem of the energy of the particle falling inside
the black hole.
    It is well known that the definition of the energy depends on
the definition of time, but time is different outside and inside of
the black hole.
    It is often claimed that space and time are somehow changed going one
to another after crossing the horizon of the black hole.
    Very popular is the idea that differently from the region outside of
the static black hole particles inside the black holes can have
negative energies.
    Usually it is discussed concerning the Hawking effect of the evaporation
of black holes due to particle creation from vacuum by the gravitation
of the hole.
    It is claimed that a pair of particles with total energy equal to zero
is created by the static gravitational field of the black hole so that
the particle with negative energy goes inside the black hole while the particle
with positive energy is going outside and is observed as Hawking radiation.
    The mass of the black hole in this process due to receiving the negative
energy becomes smaller so that one can speak about the gradual annihilation
of the black hole itself.

    In Ref.~\refcite{Hawking75}, \, S. Hawking wrote:
    ``Just outside the event horizon there will be virtual pairs of particles,
one with negative energy and one with positive energy.
    The negative particle is in a region which is classically forbidden but
it can tunnel through the event horizon to the region inside the black hole
where the Killing vector which represents time translations is spacelike.
    In this region the particle can exist as a real particle with a timelike
momentum vector even though its energy relative to infinity as measured by the
time translation Killing vector is negative.
    The other particle of the pair, having a positive energy, can escape
to infinity where it constitutes a part of the thermal emission ...
    The probability of the negative energy particle tunneling
through the horizon is governed by the surface gravity      
since this quantity measures the gradient of the magnitude of the Killing
vector or, in other words, how fast the Killing vector is becoming spacelike.
    Instead of thinking of negative energy particles tunneling through
the horizon in the positive sense of time one could regard them as positive
energy particles crossing the horizon on pastdirected world-lines and then
being scattered on to future-directed world-lines by the gravitational field.''

    Surely one can obtain the Hawking effect without this appeal to negative
energies considering the collapse of the star and particle creation
in this case when going to the static limit of the black hole.\cite{GMM}
    However let us try to investigate the problem: in what sense one can
speak about the existence of particles with negative energies inside
the Schwarzschild black hole?
    If this energy has a large absolute value can such a black hole become
a wormhole?
    It is known that negative energies are needed for the wormhole formation
and it is very attractive to think about the physical black holes as wormholes.

\vspace{14pt}
{\section{The Energy Conserved Relative to Infinity}
\label{4secUFN}
}
\noindent
    Let the spacetime have the timelike Killing vector~$\zeta^i$
orthogonal to some set of spacelike hypersurfaces~\{$\Sigma$\}.
    From the definition of the Killing vector one has
    \begin{equation} \label{NEnergy1}
\nabla^i \zeta^k + \nabla^k \zeta^i = 0 \,.
\end{equation}
    Let $T_{ik}$ be the metrical energy-momentum tensor of some field or
covariantly conserved energy-momentum tensor of some matter.
    The translational symmetry with the generator~$\zeta^i$
leads to the conservation of the quantity
    \begin{equation} \label{NEnergy2}
E^{(\zeta)} = \int_\Sigma T_{ik}\, \zeta^i \, d \sigma^k ,
\end{equation}
     which follows from the covariant conservation of~$T_{ik}$
and Eq.~(\ref{NEnergy1}) leading to~$ \nabla^i (T_{ik} \zeta^k)=0 $.
    The quantity~$ E^{(\zeta)} $ plays the role of the energy.

    If the vector~$\zeta^i$  is not timelike but it is still the Killing
vector, while~\{$\Sigma$\} is some set of hypersurfaces not necessarily
spacelike then formula~(\ref{NEnergy2}) is still defining some conserved
quantity.

    If~$\zeta^i$ is not a Killing vector then the quantity defined
by~(\ref{NEnergy2}) generally is not conserved.
    In part~\ref{5secUFN} we shall consider~(\ref{NEnergy2}) also for
such situations.

    The action for a classical pointlike particle with mass~$m$ is equal to
    \begin{equation} \label{NEnergy4}
S = - mc \int \! ds\,,
\end{equation}
    where $c$ is the light velocity.
    Let us find its energy-momentum tensor.
    Due to the definition one has
    \begin{equation}
T_{ik}= \frac{2 c}{\sqrt{ |g|}} \, \frac{\delta S}{\delta g^{ik}}
\ \ \ \Leftrightarrow \ \ \
T^{ik}= - \frac{2 c}{\sqrt{ |g|}} \, \frac{\delta S}{\delta g_{ik}} \,.
\label{dopTEMdop2}
\end{equation}
    Here $g_{ik}$ is a spacetime metric, $g={\rm det}\,\{g_{ik}\} $
and we take into account the relation
$ g^{ik} \delta g_{lk} = - g_{lk} \delta g^{ik} $. \
    From~(\ref{dopTEMdop2}) by variation of the action~(\ref{NEnergy4})
one obtains (see Ref.~\refcite{Weinberg}, Chap.~12, Sec.~2)
the energy-momentum tensor of the classical pointlike particle of mass~$m$,
at the point with coordinates~$x_{p}$ as
    \begin{equation} \label{NEnergy5}
T^{ik}(x) = \frac{m c^2 }{\sqrt{|g|}} \int \! ds \, \frac{d x^i}{d s}\,
\frac{d x^k}{d s}\, \delta^4 ( x - x_{p})\,.
\end{equation}

    The value of~(\ref{NEnergy2}) for a pointlike particle in a general metric
and for an arbitrary vector~$ \zeta^i $ is
\vspace{-10pt}
    \begin{equation} \label{NEnergyEgen}
E^{(\zeta)} = m c^2\, \frac{dx^i}{ds}\,  g_{ik} \zeta^k =
m c^2 \, (u, \zeta) = c (p, \zeta) \,,
\end{equation}
    where $ u^i= dx^i / ds $ is the four velocity, \,
$ p^i= m\, c\, dx^i / ds $ is the four momentum.

    If $\zeta^i$ is the Killing vector, then $E^{(\zeta)}$ is conserved
for a pointlike particle with the action~(\ref{NEnergy4}),
i.e. for a particle freely moving on a geodesic line.
    This also follows from~(\ref{NEnergyEgen}),
because the scalar product $(u, \zeta)$
is constant on the geodesic (see problem 10.10 in Ref.~\refcite{LPPT}).

    If $ \zeta^i $ is the translation vector on $x^0$ then
    \begin{equation} \label{NEnergy3}
\zeta^i = (1,\,0,\,0,\,0)
\end{equation}
     and~(\ref{NEnergy2}) for the pointlike particle is
    \begin{equation} \label{NEnergyE}
E^{(\zeta)} = m c^2 g_{0i}\, \frac{dx^i}{ds} = m c^2 u_0\,.
\end{equation}

    For a massless particle the energy at infinity can be defined by
a formula analogous to~(\ref{NEnergyEgen}).
    Let $\tau$ be some affine parameter on the geodesic (for a timelike
geodesic the role of this parameter plays the proper time).
    Then  for a massless particle (photon) define
    \begin{equation} \label{NEnergyEgenF}
E^{(\zeta)} = E_\infty \, \frac{dx^i}{c\, d \tau}\,  g_{ik} \zeta^k =
E_\infty \, (u, \zeta) \,,
\end{equation}
    where $ u^i= dx^i / (c\, d \tau) $, $E_\infty$  is some constant with
the dimension of energy
($E_\infty = h \nu$ for the photon, if the parameter $\tau = t$ at infinity,
$\nu$ is the frequency at infinity).

    The metric of the static uncharged black hole in Schwarzschild
coordinates has the form
    \begin{equation} \label{UFNSch}
ds^2=\left(1-\frac{r_g}{r} \right) c^2 dt^2 - \frac{dr^2}{
 \displaystyle 1-\frac{\mathstrut r_g}{r}} -
r^2 \left( d \vartheta^2 +\sin^2\! \vartheta \, d \varphi^2 \right).
\end{equation}
    Here $r_g= 2 G M / c^2 $ is the gravitational radius of the black hole
of mass~$ M $.
    For a Schwarzschild metric~(\ref{UFNSch}) vector~(\ref{NEnergy3})
is the timelike Killing vector and one gets from~(\ref{NEnergyE})
    \begin{equation} \label{NEnergy6}
E^{(\zeta)} = m c^2 \left( 1 - \frac{r_g}{r} \right) \frac{dx^0}{ds}
= m c^2 \left( 1 - \frac{r_g}{r} \right) \frac{d t}{d \tau}\,,
\end{equation}
    where $ d \tau = ds /c $.

    Outside of the horizon~(\ref{NEnergy6}) has the meaning of
the energy of the particle in this metric
    \begin{equation} \label{NEnergy7}
E = m c^2 \sqrt{ \frac{\displaystyle { 1 - \frac{r_g}{r} }}
{\displaystyle {1 - \frac{{\bf v}^2}{c^2} }} }\,,
\end{equation}
    where ${\bf v}$ is the velocity measured by the observer at rest in
the Schwarzschild coordinates.
    Inside the horizon ($r<r_g$) the expression~(\ref{NEnergy6})
also gives the value of some conserved entity called the energy
relative to infinity.
    However in this case the Killing vector~(\ref{NEnergy3})
is spacelike, the coordinate~$t$ is not timelike but spacelike and
formula~(\ref{NEnergy7}) is incorrect.

\begin{figure}[ht]
\includegraphics[height=63mm]{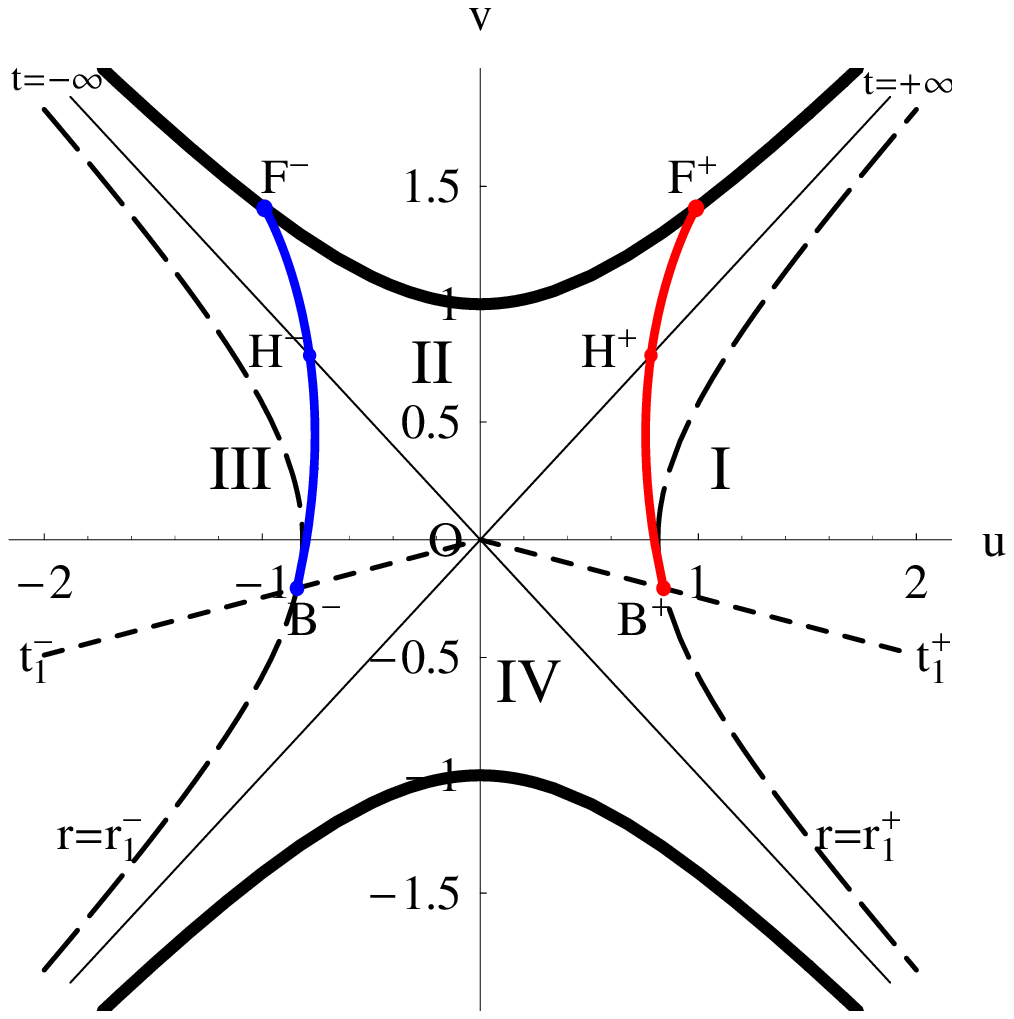}
\includegraphics[height=63mm]{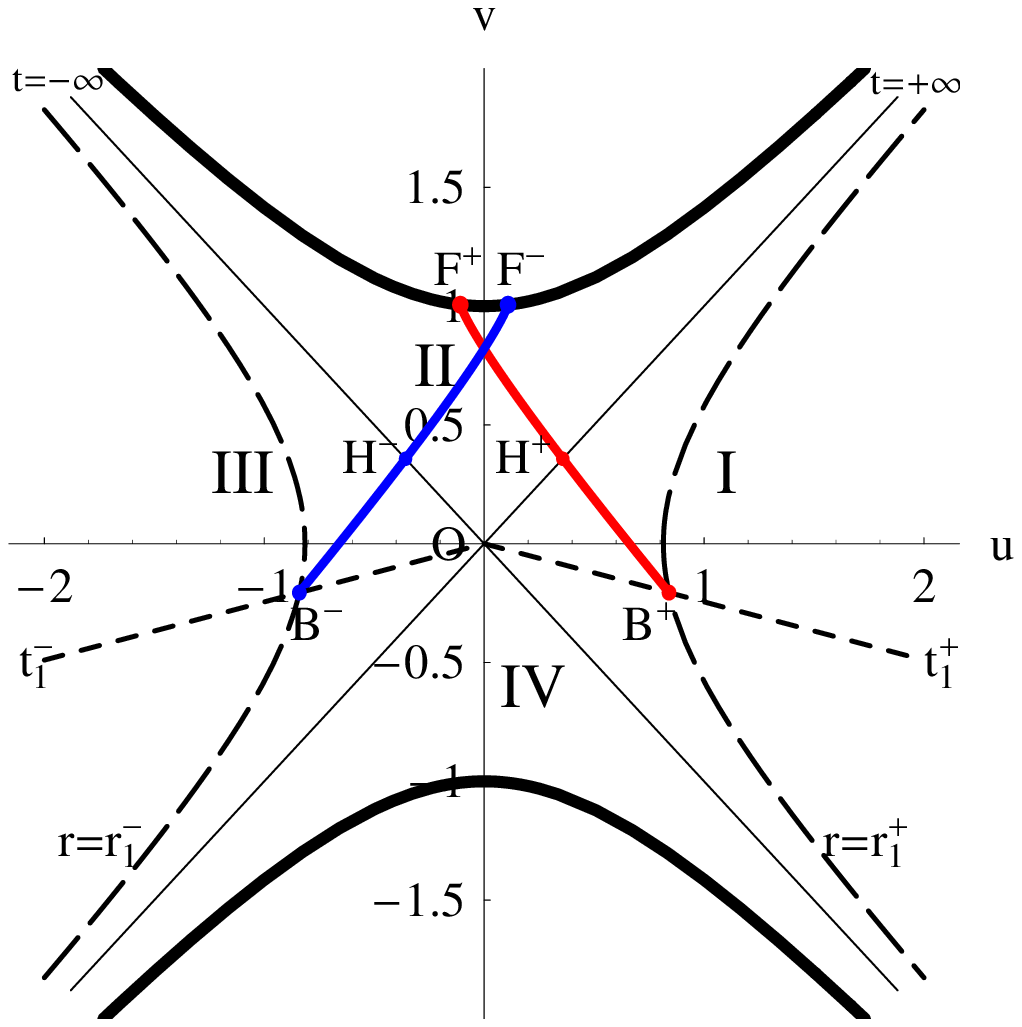}
\caption{\small
The radial  falling inside the black hole of the massive particles
with positive ($B^+H^+F^+$) and negative ($B^-H^-F^-$) energy.
Here $t_1^+=-0.5 r_g/c, \ t_1^-=+0.5 r_g/c$, \
$r_1^+ = r_1^- = 1.2 r_g$,
the energy is $|E|/mc^2=0.41 $         
in the left figure and $0.7 $ in the right one.}
\label{NegEnergy}
\end{figure}

    Negative value of the energy outside the horizon
due to~(\ref{NEnergy6}) is possible either for particles
with negative mass or for particles with positive mass but moving
backwards in time.
    Both cases do not have clear physical sense.
    However if the energy of the particle with positive mass inside
the horizon ($r<r_g$) is negative relative to infinity this means
due to~(\ref{NEnergy6}) that $dt / d \tau > 0 $.
    {\sl Inside the horizon particles with $m>0$ are moving to the right
along the space coordinate~$t$ if their energy relative to infinity is
negative and they are moving to the left if their energy is positive}!
    The examples of movement of particles with positive and negative energies
are represented in Fig.~\ref{NegEnergy} in Kruskal-Szekeres coordinates.

    The analogous situation exists for massless particles (photons).
    If their energy is positive relative to infinity, then inside the horizon
$dt / d \tau < 0 $ represents movement to the left along an angle of $45^\circ$
on Kruskal-Szekeres diagram.
    If their energy is negative relative to infinity, then inside the horizon
$dt / d \tau > 0 $ represents movement to the right along an angle
of $45^\circ$.

    The analysis of the light cones shows that for particles crossing
the event horizon movement to the left in~$t$ corresponds for
the particle falling inside the black hole in Kruskal--Szekeres coordinates
to movement (at the moment of crossing the horizon) in the region  {\bf II}
in the first quadrant, while movement to the right in~$t$ (at the same moment)
means movement in the region {\bf II} but in the second quadrant.
    The Kruskal--Szekeres diagram for the star collapsing into a black hole
is given in Ref.~\refcite{MTW}, Fig.~32.1.    
    In this case the region {\bf II} in the second quadrant is absent.
    So the interpretation of the process of radiation as creation of particles
with negative energy exactly on the horizon is impossible!

\vspace{4pt}
    Expression~(\ref{NEnergy6}) can be obtained in another way.
    As it is known the geodesics in spacetime with
metric~$d s^2 = g_{ik} dx^i dx^k$
can be obtained (see Ref.~\refcite{Chandrasekhar}, Chap.\,1, Sec.~6) as
the Euler-Lagrange equations for the extremal problem for the functional
    \begin{equation} \label{NEnergy8}
S = \frac{1}{c} \int \! L\, ds \,, \ \ \ \ \
{L} = \frac{mc^2}{2}\, g_{ik}\, \frac{dx^i}{ds} \frac{d x^k}{ds}\,.
\end{equation}

    If the Lagrangian~(\ref{NEnergy8}) does not explicitly depend on
the variable~$t$, then the corresponding momentum is conserved:
    \begin{equation} \label{NEnergy10}
p_t  = m c^2 g_{0i}\, \frac{d x^i}{d s} = {\rm const}\,,
\end{equation}
    which coincides due to the formula~(\ref{NEnergyE}) with the energy
of the particle $ E^{(\zeta)} $.

    For example,
in Schwarzschild spacetime~(\ref{UFNSch}) the Lagrangian~$L$ has the form
    \begin{equation} \label{NEnergy9}
L = \frac{mc^2}{2}\, \left[ \left(1-\frac{r_g}{r} \right) {\dot
t}^2- \frac{{\dot r}^2}{\displaystyle c^2 \! \left(1-\frac{r_g}{r} \right) } -
\frac{ r^2}{c^2} \left( {\dot \vartheta}^2 +
\sin^2\! \vartheta \, {\dot \varphi}^2 \right) \right],
\end{equation}
    where the dot denotes derivative with respect to the proper time $ \tau $.
    The momentum $p_t$ is equal to~(\ref{NEnergy6}) and it is conserved
because the Lagrangian~(\ref{NEnergy9}) does not depend explicitly on
the variable~$t$.
    So the conserved momentum~$p_t$ is identical to the energy of the particle
relative to infinity in Schwarzschild gravitational field.

\vspace{17pt}
{\section{The Nonconserved Energy Inside the Black Hole}
\label{5secUFN}
}

{\subsection{The Schwarzschild coordinates inside the black hole}
\label{OtrEnsec31}
}
\noindent
    Using  notations
    \begin{equation} \label{OtrEn31sec}
\eta=- \frac{r}{c}\,, \ \ \eta \in \left(- \frac{r_g}{c}\,,\, 0 \right);
\ \ \ \ l=ct \,, \ \ l \in {\rm \mathbf{R}}\,,
\end{equation}
    the metric inside the horizon is given by
    \begin{equation} \label{UFNScwIInlDop}
d s^2 = \frac{c^2 d \eta^2}{\displaystyle \biggl(\,
\frac{\mathstrut  r_g}{- c \eta} -1 \biggr)}
- \left( \frac{r_g}{- c \eta}-1 \right) d l^2 -
(c \eta)^2 \left( d \vartheta^2 +\sin^2\! \vartheta \, d \varphi^2 \right).
\end{equation}
    From~(\ref{UFNScwIInlDop}) it is seen that the coordinate~$\eta$
plays the role of time inside the black hole.
    The negative sign in the first equation of~(\ref{OtrEn31sec}) is chosen
so that the future direction inside the black hole corresponds to
increasing~$\eta$.
    Choose coordinates
$x^0 = c \eta , \ x^1 = l , \ x^2 = \vartheta , \ x^3 =\varphi$.
    Then the 4-vector~(\ref{NEnergy3})
is the generator of translations in time~$\eta$
(here it is not a Killing vector).
    The energy relative to such translations defined by~(\ref{NEnergy2})
surely is not conserved.
    From~(\ref{NEnergyE}) one obtains
    \begin{equation} \label{NEnergy12}
E^{(\eta)} = \frac{\displaystyle m c^2\, \frac{d \eta}{d \tau} }
{\displaystyle \left( \frac{r_g}{-c \eta} - 1 \right)}
= - \frac{\displaystyle  m c \, \frac{d r}{d \tau} }
{\displaystyle  \left( \frac{r_g}{r} - 1 \right)}.
\end{equation}
    For movement inside the black hole one has $dr/ d \tau < 0$
 and $ E^{(\eta)} > 0 $\,.
    Let us find more exact limitations on the possible values of~$ E^{(\eta)} $
for a massive particle inside the black hole.
    From the definition of the four velocity one has $u^i u_i =1$ and
for the Schwarzschild metric one gets
    \begin{equation} \label{uiuiShwar}
\left(1-\frac{r_g}{r} \right)\! \left( \frac{c dt}{d s} \right)^{\!2} -
\frac{1}{ \displaystyle 1-\frac{\mathstrut r_g}{r}}
\left( \frac{dr}{d s} \right)^{\!2} -
r^2 \left( \! \left( \frac{d\vartheta}{d s} \right)^{\!2} +
\sin^2\! \vartheta \left( \frac{d \varphi}{d s} \right)^{\!2} \right) = 1.
\end{equation}
    Inside the horizon $r< r_g$ all terms in the left side of~(\ref{uiuiShwar})
are negative except the term with $dr/ds$, so this term is larger or equal
to one and
    \begin{equation} \label{uiuiShwar2}
- \frac{dr}{d \tau} \ge c \, \sqrt{ \frac{r_g}{r} - 1 }\,, \ \ \ \ r < r_g .
\end{equation}
    So possible values of the ``energy''~$ E^{(\eta)}(r) $ for a given~$r$
inside the black hole satisfy the inequality
    \begin{equation} \label{uiuiShwar3}
E^{(\eta)}(r) \ge \frac{ m c^2 }{\displaystyle \sqrt{ \frac{r_g}{r} - 1 }}\,,
\ \ \ \ r < r_g .
\end{equation}

    The geodesics (with $\vartheta=\pi/2$) in Schwarzschild metric satisfy
the equations
    \begin{equation} \label{UFNSGdL}
\left( \frac{d r}{c\, d \tau} \right)^{\!2} =
\Bigl( \frac{r_g}{r} -1 \Bigr) \Bigl(1 + \frac{L^2}{r^2} \Bigr)
+ \varepsilon^2, \ \ \ \
r^2 \frac{d \varphi}{c\, d \tau} =L={\rm const}
\end{equation}
    (see Ref.~\refcite{Chandrasekhar}, Sec.~19),    
where $\varepsilon={\rm const}$ is the specific energy relative to infinity,
$L$ denotes the angular momentum about an axis normal to the invariant
plane in units of~$mc$.
    So the energy~$ E^{(\eta)} $ introduced in~(\ref{NEnergy12})
for the freely falling particle with mass~$m$ is
\vspace{-5pt}
    \begin{equation} \label{NEnergy13}
E^{(\eta)}(r) = \frac{m c^2}{\frac{\textstyle \mathstrut r_g}
{\textstyle \mathstrut r} -1}\,
\sqrt{ \varepsilon^2 + \Bigl( \frac{r_g}{r} - 1 \Bigr)
\Bigl(1 + \frac{L^2}{r^2} \Bigr)}
\end{equation}
    and it changes from $+ \infty$ on the horizon to $0$ in the singularity,
if $L=0$.
    For geodesics with angular momentum $ L \ne 0$,
the energy $ E^{(\eta)} $ inside the horizon changes from $+ \infty$
on the horizon, decreases to some finite positive value and then
it is again increasing to $ + \infty$  in the singularity.

    Outside the hole the value~(\ref{NEnergy12}) is not an energy
and can be positive or negative according to the sign of $d r / d \tau $.

\vspace{7pt}
{\subsection{The interior of the black hole as anisotropic cosmology}
\label{OtrEnsec32}
}
\noindent
    The spacetime inside the Schwarzschild black hole is
an example of homogeneous space but anisotropic cosmological
model.\cite{KantowskiSachs66}
    Really, if one introduces the variable  
    \begin{equation} \label{UFNScwMet3}
d \xi = d \eta  \sqrt{ \frac{- c \eta}{r_g  + c \eta}}, \ \ \
\xi = \sqrt{- \eta \Bigl( \frac{ r_g}{c} + \eta \Bigr) }
- \frac{r_g}{c} \tan^{-1} \sqrt{
\frac{- c \eta}{r_g  + c \eta}} , \ \ \
\xi \in \left( - \frac{ \pi r_g}{2 c}, \, 0 \right)\!,
\end{equation}
    the metric inside the horizon can be written as
    \begin{equation} \label{UFNScwMet2}
d s^2 = c^2 d \xi ^2 - f_1 (\xi)\, d l^2 - f_2 (\xi)
\left( d \vartheta^2 +\sin^2\! \vartheta \, d \varphi^2 \right),
\end{equation}
    where $f_1 (\xi)$ and $ f_2 (\xi)$ are some functions which can be found
from~(\ref{UFNScwIInlDop}) and~(\ref{UFNScwMet3}).

    The energy~(\ref{NEnergyE}) relative to  translations in time~$\xi$
is equal to
    \begin{equation} \label{NEnergy15}
E^{(\xi)} = m c^2\, \frac{d \xi}{d \tau}
= - \frac{\displaystyle m c\, \frac{d r}{d \tau} }
{\displaystyle  \sqrt{ \frac{r_g}{r} - 1 }}\,.
\end{equation}
    Inside the black hole one has the inequality~(\ref{uiuiShwar2}) and so
    \begin{equation} \label{NEnergymc2}
E^{(\xi)}(r) \ge m c^2, \ \ \ \ r < r_g.
\end{equation}
    Using~(\ref{UFNSGdL}) one can see that for a particle moving along
a geodesic inside the black hole, the energy $ E^{(\xi)} $ thus defined
is equal to
    \begin{equation} \label{NEnergy16}
E^{(\xi)}(r) = m c^2\,
\sqrt{ \frac{ \varepsilon^2 r }{ r_g - r } + 1 + \frac{L^2}{r^2}} \,.
\end{equation}
    For a radially  falling particle $E^{(\xi)}$ is decreasing from $+\infty$
on the horizon to $ mc^2 $ at the singularity.
    For geodesics with nonzero angular momentum the energy $ E^{(\xi)} $
inside the horizon changes from $+ \infty$ on the horizon, decreases to
some positive value and then again increases to $ + \infty$ at the singularity.

    Outside the horizon the variable~$\xi$ and the energy~$E^{(\xi)}$
have imaginary values.

\vspace{7pt}
{\subsection{Energy inside the black hole for conformal time}
\label{OtrEnsec33}
}
\noindent
    If one introduces in~(\ref{UFNScwIInlDop}) the variable
    \begin{equation} \label{UFNScw19}
d \tilde{\xi} = \frac{- c \eta\, d \eta}{r_g + c \eta}, \ \ \
\tilde{\xi} = - \eta + \frac{r_g}{c} \ln \left( 1 + \frac{c \eta}{r_g} \right),
\ \ \ \tilde{\xi} \in \left( - \infty, \, 0 \right)\!,
\end{equation}
    the metric inside the horizon can be written as
    \begin{equation} \label{UFNScw18}
d s^2 = a^2(\tilde{\xi} )\left( c^2 d {\tilde{\xi}}^{\,2} - d l^2 -
f(\tilde \xi)
\left( d \vartheta^2 +\sin^2\! \vartheta \, d \varphi^2 \right) \right),
\end{equation}
    where $a(\tilde{\xi})$ and $ f(\tilde{\xi})$  are some functions which
can be obtained from~(\ref{UFNScwIInlDop}) and~(\ref{UFNScw19}) and
$\tilde \xi$ is the analog of the conformal time in homogeneous isotropic
cosmological models.

    The energy relative to translations in time~$ \tilde{\xi} $ is equal
to~$ E^{(\tilde \xi)} = - m c\, d r / d \tau $
and due to~(\ref{uiuiShwar2}) it satisfies the limitation
    \begin{equation} \label{NEnergytxi}
E^{(\tilde \xi)}(r) = - m c \, \frac{d r}{d \tau} \ge
m c^2 \, \sqrt{ \frac{r_g}{r} - 1 }\,, \ \ \ \ r < r_g.
\end{equation}
    For the particle falling on the geodesic~(\ref{UFNSGdL})
    \begin{equation} \label{NEnergy20}
E^{(\tilde \xi)} = - m c \frac{d r}{d \tau} =
m c^2 \sqrt{ \left( \frac{r_g}{r} - 1 \right)
\left( 1 + \frac{l_z^2}{r^2} \right) + \varepsilon^2 },
\end{equation}
 $E^{(\tilde \xi)}$ is growing
 from $ mc^2 |\varepsilon|$ on the horizon to $+\infty$  at the singularity.

   So our examples show that for particles falling inside the black hole one
can introduce differently some value analogous to the energy which
is non negative but it is not conserved in time.
    This situation is analogous to that in nonstationary external field.
    The gravitational field inside the Schwarzschild black hole surely
is not static.

\vspace{17pt}
{\section{Kerr's Metric}
\label{6secUFN}
}
\noindent
    Let us compare the situation with negative energy inside the static
black hole with the well known case of the rotating black hole.
Kerr's metric in Boyer--Lindquist coordinates is
    \begin{eqnarray}
d s^2 = d t^2 - (r^2 + a^2) \sin^2 \! \theta\, d \varphi^2 -
\frac{2 M r \, ( d t - a \sin^2 \! \theta\, d \varphi )^2}{r^2 + a^2 \cos^2
\! \theta } -    \nonumber \\
-\, (r^2 + a^2 \cos^2 \! \theta ) \left( \frac{d r^2}{r^2 - 2 M r + a^2} +
d \theta^2 \right),
\label{Kerr}
\end{eqnarray}
    where $M$ is the mass of the black hole, $aM$ its angular momentum.
    Here we use units where $G=c=1$.
    On the event horizon one has
     \begin{equation}
r = r_H \equiv M + \sqrt{M^2 - a^2} .
\label{Hor}
\end{equation}
    The surface called the static limit is defined by
     \begin{equation}
r = r_0 \equiv M + \sqrt{M^2 - a^2 \cos^2 \! \theta} .
\label{PrSt}
\end{equation}
    The spacetime region between the horizon and the static limit is called
ergosphere.
    It is in the ergosphere that  one has the trajectories of particles with
negative energy.
    Inside the ergosphere the Killing vector of translations in
time $(1,0,0,0)$
    becomes spacelike similar to that discussed by S.\,Hawking for
static black holes in the citation in the beginning of this paper.
    The energy of the particle relative to infinity can be negative.

    Really in the case of the spacelike Killing vector $(k_0, \mathbf{ k}) $
in some reference frame (see problem 10.15 in Ref.~\refcite{LPPT})
one can write
    \begin{equation}
E^{(k)} = m \gamma (k_0 - \mathbf{ v} \cdot \mathbf{ k}),
\label{KerrEnergySP}
\end{equation}
    where $\gamma = 1/ \sqrt{1 - \mathbf{v}^2} $, \,
$\mathbf{v}$ is the three-velocity of a particle,
$ k_0 < |\mathbf{ k}|$ and $ E^{(k)} <0 $ is possible.

    Due to formula~(\ref{NEnergyE}) one gets
    \begin{equation}
E^{(\zeta)} = m \left[ \left( 1 - \frac{2 M r}{r^2 + a^2 \cos^2 \theta} \right)
\frac{d t}{d \tau} + \frac{2 M r a \sin^2 \theta}{r^2 + a^2 \cos^2 \theta} \,
\frac{d \varphi}{d \tau} \right].
\label{KerrEnergy}
\end{equation}
    $ E^{(\zeta)} <0 $ is possible for $r< 2 M$ and movement of particles
rotating in the direction opposite to rotation of the black hole
(see Ref.~\refcite{Chandrasekhar}, Sec.~65).    
    Existence of states with negative energy in the ergosphere leads to the
possibility of getting energy from the rotating black hole.\cite{Penrose69}

    In all the cases considered, the states with negative relative to infinity
energy exist for the spacelike Killing vector either in ergosphere of the
Kerr's black hole or inside the horizon of the Schwarzschild's black hole.

    If one has other than gravitational interaction of the pointlike particle
then states with conserved energy are possible for the timelike Killing vector.
    Let us consider this situation for the example of the charged
nonrotating black hole.

\vspace{39pt}
{\section{Reissner--Nordstr{\o}m Black Holes}
\label{secReisNord}
}
\noindent
    The Reissner--Nordstr{\o}m solution for the metric of the static
charged black hole has the form
    \begin{equation} \label{ResnNordstrom}
ds^2= \frac{\Delta}{r^2}\, dt^2 - \frac{r^2}{\Delta}\,dr^2
- r^2 \left( d \vartheta^2 +\sin^2\! \vartheta \, d \varphi^2 \right), \ \ \
\Delta = r^2 - 2 M r + Q^2,
\end{equation}
    where $Q$ is the charge of the black hole with mass~$M$
(here we used the system of units $G=c=1$).
    The roots of equation~$\Delta =0$,
    \begin{equation} \label{ResnNordHor}
r_H = M +\sqrt{M^2 - Q^2}, \ \ \ \ r_C = M - \sqrt{M^2 - Q^2},
\end{equation}
    define the surfaces which are the event horizon $r=r_H$ and the
Cauchy horizon $r=r_C$ for the charged black hole.

    Equations of movement of the particle with specific (divided by mass)
charge~$q$ in the metric~(\ref{ResnNordstrom}) are
(see Ref.~\refcite{Chandrasekhar}, Sec.~40)
    \begin{equation} \label{ResnNordGeod}
\frac{\Delta}{r^2} \frac{d t}{d \tau} + \frac{q Q}{r} = \varepsilon =
{\rm const}, \ \ \ \ r^2 \frac{d \varphi}{d \tau} = L ={\rm const},
\end{equation}
    \begin{equation} \label{ResnNordGeod2}
\left( \frac{d r}{d \tau} \right)^2 + \frac{\Delta}{r^2}
\left( 1 + \frac{L^2}{r^2} \right) =
\left( \varepsilon - \frac{q Q}{r} \right)^2.
\end{equation}
    As one can see from~(\ref{ResnNordGeod}), (\ref{ResnNordGeod2})
for $q Q <0$ one has states of the charged particle with negative value
of the specific energy~$\varepsilon \approx q Q / r_H $ in the vicinity
of the event horizon~$r_H$.
    Existence of states with negative energy leads to the possibility
of getting energy from the charged
black hole.\cite{ChristodoulouRuffini71,DenardoRuffini72}

    For noncharged particles (with positive mass) outside the event horizon
states with negative energy don't exist.
    From~(\ref{ResnNordGeod}) one comes to the conclusion that if only
movement to the future in time is existing, i.e. $dt / d \tau >0$
then $q=0$ leads to $\varepsilon >0$.

    So negative relative to infinity energies are possible also for the
timelike Killing vector but for the nongeodesic particle movement which
occurs for the electromagnetic interaction of the charged particle in the
field of the Reissner--Nordstr{\o}m black hole.
    Negative relative to infinity energies in the given metric in all cases
lead to the possibility of extracting energy from the given object.
    If these negative energies exist inside the event horizon the extraction
of energy is possible due to the quantum process --- the Hawking effect.
    If negative energies exist outside the event horizon extraction of energy
occurs due to the classical process (Penrose process in the ergosphere of
the rotating black hole).

\vspace{14pt}


\begin{thebibliography}{00}

\bibitem{Castagnino89}
M. Castagnino, {\it Phys. Rev.~D} {\bf 39} (1989) 2216.

\bibitem{Castagnino98}    
M. Castagnino, {\it Phys. Rev.~D} {\bf 57} (1998) 750.

\bibitem{Hawking75}
S. W. Hawking,
{\it Commun. Math. Phys.} {\bf 43} (1975) 199.

\bibitem{GMM}
A. A. Grib, S. G. Mamayev and V. M. Mostepanenko,
{\it Vacuum Quantum Effects in Strong Fields}
(Friedmann Lab. Publ., St.\,Petersburg, 1994).

\bibitem{Weinberg}    
S. Weinberg, {\it Gravitation and Cosmology}
(Wiley, New York, 1972).

\bibitem{LPPT}
A. P. Lightman, W. H. Press, R. H. Price and S. A. Teukolsky,
{\it Problem Book in Relativity and Gravitation\/}
(Princeton Univ. Press, New Jersey, 1975).

\bibitem{Chandrasekhar}
S. Chandrasekhar, {\it The Mathematical Theory of Black Holes}
(Clarendon Press, Oxford, 1983).

\bibitem{MTW}
C. W. Misner, K. S. Thorne and J. A. Wheeler,
{\it Gravitation\/} (W. H. Freeman, San Francisco, 1973).

\bibitem{KantowskiSachs66}
R. Kantowski and R. K. Sachs,
{\it J. Math. Phys.} {\bf 7} (1966) 443.

\bibitem{Penrose69}
R. Penrose, {\it Rivista Nuovo Cimento} {\bf I}, Num. Spec., (1969) 252.

\bibitem{ChristodoulouRuffini71}
D. Christodoulou and R. Ruffini,
{\it Phys. Rev.~D} {\bf 4} (1971) 3552.  

\bibitem{DenardoRuffini72}
G. Denardo and R. Ruffini,
On the energetics of Reisner Nordstr{\o}m geometries,
{in} {\it Black Holes},
eds. C. DeWitt and B. S. DeWitt
(Gordon and Breach, New York, 1973), p.~R33.    

\end{thebibliography}
\end{document}